\documentclass[preprint2]{aastex} 
\usepackage{amssymb, amsmath, graphics} 
\usepackage{longtable,graphicx} 
\usepackage{aas_macros} 
 
\usepackage[usenames]{color}

\shorttitle{Detection of mode coupling} 
\shortauthors{Michel Breger et al.}
 
\begin{document}

\title{Evidence of resonant mode coupling and the relationship between low and high frequencies in a rapidly rotating A star}

\author
{M.~Breger\altaffilmark{1,2}, M.~H.~Montgomery\altaffilmark{1}
\altaffiltext{1}{Department of Astronomy, University of Texas, Austin, TX 78712, USA}
\altaffiltext{2}{Institut f\"ur Astrophysik der Universit\"at Wien, T\"urkenschanzstr. 17, A--1180, Wien, Austria}
}  
\begin{abstract} 

  In the theory of resonant mode coupling, the parent and child modes are directly related in frequency and phase.  The oscillations present in the fast rotating $\delta$\,Sct star KIC\,8054146 allow us to test the most general and generic aspects of such a theory. The only direct way to separate the parent and coupled (child) modes is to examine the correlations in amplitude variability between the different frequencies.  For the dominant family of related frequencies, only a single mode and a triplet are the origins of nine dominant frequency peaks ranging from 2.93 to 66.30 cycles day$^{-1}$ (as well as dozens of small-amplitude combination modes and a predicted and detected third high-frequency triplet).  The mode-coupling model correctly predicts the large amplitude variations of the coupled modes as a product of the amplitudes of the parent modes, while the phase changes are also correctly modeled. This differs from the behavior of ``normal'' combination frequencies in that the amplitudes are \emph{three orders of magnitude} larger and may exceed even the amplitudes of the parent modes.  We show that two dominant low frequencies at 5.86 and 2.93 cycles day$^{-1}$ in the gravity-mode region are not harmonics of each other, and their properties follow those of the almost equidistant high-frequency triplet. We note that the previously puzzling situation of finding two strong peaks in the low-frequency region related by nearly a factor of two in frequency has been seen in other $\delta$ Sct stars as well.

\end{abstract} 
 
\keywords{Stars: oscillations (including pulsations) -- Stars: variables: delta Scuti -- Stars: rotation -- Stars: individual: KIC\,8054146 -- {\it Kepler}}

\section{Introduction}

\begin{figure*}[!t]
\centering
\includegraphics[bb=40 70 750 550, width=\textwidth, clip]{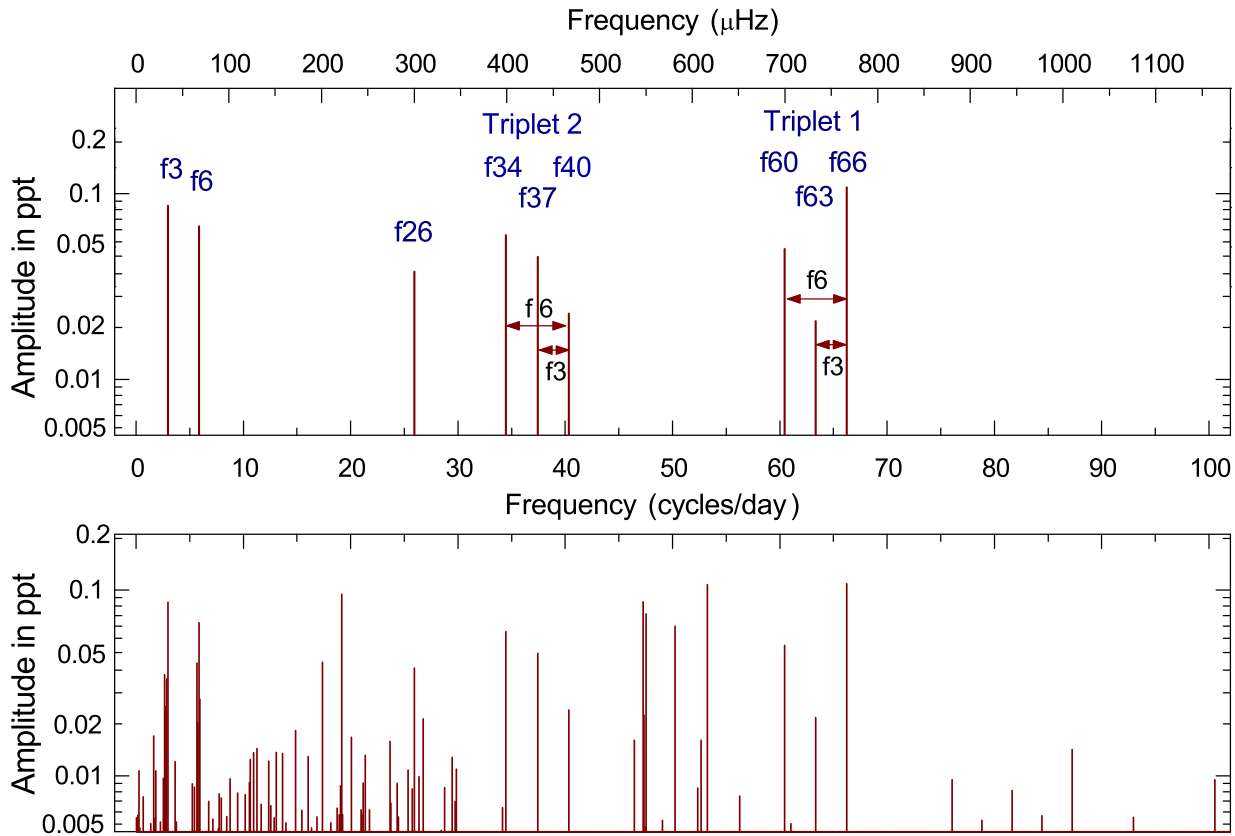}
\caption{The dominant 9 frequencies of the T family of frequencies. The frequency numbers, e.g. f26, denote their approximate values in cycles day$^{-1}$ and this notation will be used in the rest of the paper. Note that the frequencies are both in the low-frequency g-mode and high-frequency p-mode regions. The bottom panel illustrates their relation to all the detected frequencies.}
\end{figure*}

The recent spaced-based studies of $\delta$\,Sct and $\gamma$\,Dor stars have revealed that most of these stars pulsate with a multitude of pulsation modes in both the low and
high-frequency regions (Uytterhoeven et al. 2011), where the gravity (g) and pressure (p) modes are found. The numerical relationships between many low and high frequencies indicate that it is not possible to treat these regions independently of each other. For example, the low-frequency region contains g-mode pulsations as well as rotation peaks (Balona 2011,
Breger et al. 2011), while the high-frequency region contains p-modes as well as high-degree g-modes
(Monnier et al. 2010, Breger et al. 2013). Regrettably, pulsation mode identifications are generally not available for these fainter
stars studied from space; consequently, frequency patterns have been the most valuable tool. In the present investigation, we add amplitude and phase variations as an additional method to determine the modal origin.

The rapidly rotating ($V \sin i = 300\,$km/s) hybrid $\delta$\,Sct/$\gamma$\,Dor pulsator KIC\,8054146 has been measured extensively by the {\it Kepler} spacecraft. Altogether, 349 statistically significant frequencies were determined from three years of short-cadence {\it Kepler} data (Breger et al. 2012, hereafter Paper 1). The amplitudes ranged from about 2 ppm (parts-per-million) to 200 ppm.  The excellent frequency resolution of three years of data for KIC\,8054146 also revealed that many frequencies are related over and beyond the expected simple harmonics and combinations. In fact, three separate families of frequencies spanning a 200 cycles day$^{-1}$ frequency range have been discovered so far. Within each family, the amplitude variations of the low-frequency members correlate with those of the high-frequency members.
The unprecedented accuracy of the {\it Kepler} allows us examine the physical origin of these families in detail. One of these families could already be interpreted as high-degree prograde Kelvin modes (Breger et al. 2013).

\begin{table}
\caption{Dominant frequencies of the ``T'' family}
\begin{tabular}{llrl}
\hline
\noalign{\smallskip}
Name & \multicolumn{2}{c}{Frequency} & Comment\\
& cycles day$^{-1}$ & $\mu$Hz \\
\noalign{\smallskip}
\hline
\noalign{\smallskip}
f3 & 2.9308 & 33.92 & Low frequency\\
f3b & 2.9334 & 33.95 & Low frequency\\
f6 & 5.8642 & 67.87 & =f3+f3b\\
f26 & 25.9509 & 300.36 \\
f34 & 34.4836 & 399.12 & Triplet 2\\
f37 & 37.4170 & 433.07 & Triplet 2\\
f40 & 40.3479 & 466.99 & Triplet 2\\ 
f60 & 60.4346 & 699.47 & Triplet 1\\
f63 & 63.3680 & 733.43 & Triplet 1\\
f66 & 66.2988 & 767.35 & Triplet 1\\
\hline
\end{tabular}
\end{table}

This paper investigates the dominant set of modes excited in KIC\,8054146, which we have called the T family. The tools for this investigation are not only the frequency values, but also their large and steady amplitude and phase changes over the three years. 

The bottom panel of Fig. 1 shows the amplitude spectrum of KIC 8054146 in the 0 to 100 cycles day$^{-1}$ range. For more observational information we refer to Paper I, which was extended to include the {\it Kepler} short-cadence data from quarters 11 to 14 (Q11 to Q14). Since the frequency resolution of the combined Q5-Q14 data exceeds the time scales of the (small) frequency variations of the star, the diagram actually shows the sum of four subsets. The top panel of Fig. 1 presents the T family of dominant modes with additional properties of these modes listed in Table 1.

While equidistant frequency spacings and a numerical relationship between the low and high frequencies are also seen in other $\delta$ Sct stars studied by $\it Kepler$, the number and sizes of such occurrences in KIC 8054146 is unusual. In addition, the continuous short-cadence coverage over three years makes a study of the correlations of amplitude variability of the different modes possible: this allows us to uniquely separate the parent and child (coupled) modes.

\section{Determining amplitude and phase\\ variations from Q5 to Q14}

The three years of available data allow us to examine the correlations of the amplitude and phase changes of the dominant frequencies in order to detect their
physical relationship and origin. The one month of Q2 data was omitted from most of the present discussion due to the large time gap and large phase changes from Q2 to Q5. This leaves the near-continuous coverage from Q5 to Q14 spanning 928 days. For this analysis we adopt a compromise between the wish to study short-term amplitude and phase changes and an excellent frequency resolution obtained only from larger time intervals: 45-day intervals were chosen. To minimize the remaining difficulties caused by frequency resolution and small aliasing effects from short time gaps in the data, a special technique was applied, which we describe below.

\begin{figure}[!t]
\centering
\includegraphics[bb=5 140 540 780, width=\columnwidth, clip]{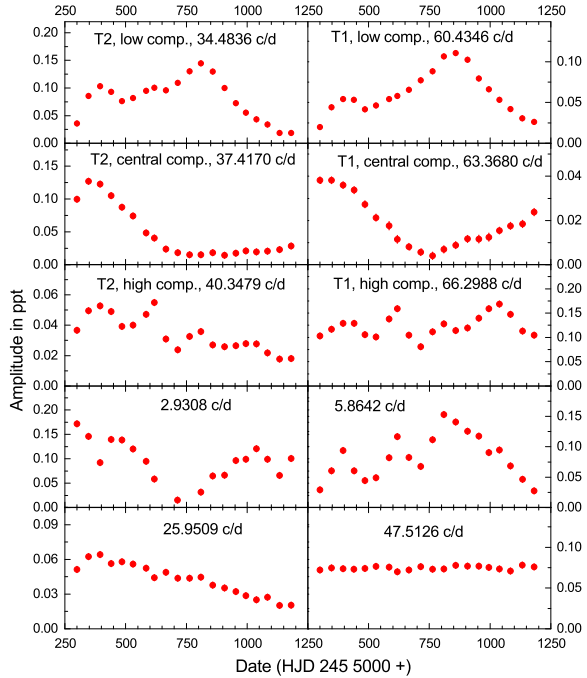}
\caption{Amplitude variations of the components of Triplet 1 (T1) and Triplet 2 (T2). Two dominant low frequencies, f26, and a mode with constant amplitudes are also shown.}
\end{figure}

\begin{figure}[!t]
\centering
\includegraphics[bb=5 140 540 780, width=\columnwidth, clip]{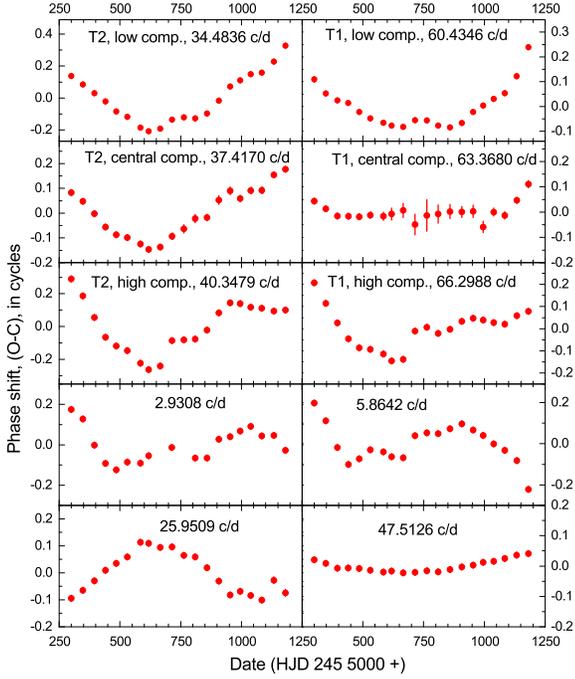}
\caption{(O-C) phase variations of the modes shown in Fig. 2. Note the qualitative similarity between the frequency changes of the related modes.}
\end{figure}

The detected frequencies were divided into two groups: the 20 dominant modes and their harmonics (for which amplitude and phase variations were desired) and the other
modes with smaller amplitudes. We divided the data into separate 270d data sets in order to minimize the effects of the amplitude and phase variations of the small-amplitude modes. We assumed constant amplitudes over 270d for these small-amplitude modes, while allowing for amplitude and phase variations of the dominant modes and their harmonics.
This allowed us to prewhiten all the small-amplitude modes. 

Two small-amplitude frequencies require a special mention:  the mode at 19.1838 cycles day$^{-1}$ showed a small amplitude of 0.01 ppt from Q5 to Q8 and slowly increased in to 0.19 ppt in amplitude from Q9 to Q14. This variation was taken into account. Another weak mode at 25.9358 cycles day$^{-1}$ forms a close double with a dominant mode at 25.9509 cycles day$^{-1}$. Because of the slow and systematic amplitude growth of this weak companion (from 0.01 to 0.03 ppt) and its separation of only 0.0151 cycles day$^{-1}$, we had to choose 45d, rather than 30d, intervals for this study.

As a next step, for each of the high-amplitude modes, average frequencies covering the entire Q5-Q13 quarters were determined. The values of the frequencies are listed in Fig. 3. Using these average frequency values, for each 45d interval the amplitudes and (O-C) shifts were calculated for the different high-amplitude modes. These statistically significant shifts show that the frequencies are slightly variable. This does not lead to problems with the adoption of constant frequency values with variable phasing, since the 45d intervals are short.

The most interesting results are connected with the dominant frequencies associated with the T family. The amplitude and (O-C) variations are shown in Figures 2 and 3. The formal error bars are also plotted; in most cases they are smaller than the symbols used. In these figures we also show the results for a mode at 47.5126 cycles day$^{-1}$, which shows very little scatter and confirms the high quality of the available data and the lack of serious systematic errors.

The figures show a very large changes in amplitude and phasing with usual time scales of a year or longer. Furthermore, these variations are steady with excellent agreement between the values of adjacent 45d time stretches. The variations vary from mode to mode, in particular for the different components of the triplets. Nevertheless, qualitative similarities can be seen between Triplets 1 and 2.

As a next step, for both the amplitude and the (O-C) variations, we have computed the correlations between the different frequencies. This led to the
discovery of a remarkably simple pattern, shown by both the amplitude and phase variations. This pattern links the triplets with each other, as well
as to the low frequencies. This will be explored in the next section.

\section{Nonlinear mode coupling}

\subsection{The simple model}

The simplest possible way that combination frequencies (i.e., ``sum and difference frequencies'') can be generated in a light curve is through a nonlinear mixing process. Examples of this are the nonlinear terms in the fluid equations and $T^4$ nonlinearity. If $x_1 = A_1\, \cos(\omega_1 t + \phi_1)$ and $x_2 = A_2\, \cos(\omega_2 t + \phi_2)$, then the lowest order nonlinear signal that will be generated is $x_3$, where \begin{eqnarray}
  x_3(t) & \propto & x_1(t) \, x_2(t) \nonumber \\
         & \propto & A_1 A_2 \cos(\omega_1 t + \phi_1) \cos(\omega_1 t + \phi_1) \nonumber \\
         & \equiv & A_{3} \cos(\omega_{3+} t + \phi_{3+}) \nonumber \\
         &    +   & A_{3} \cos(\omega_{3-} t + \phi_{3-}), 
\end{eqnarray}
where, 
\begin{eqnarray}
A_3 & = & \frac{1}{2} A_1 A_2 \label{amp} \\
\omega_{3\pm} & = & \omega_1\pm\omega_2 \label{freq} \\
\phi_{3\pm} & = &\phi_1\pm\phi_2 \label{phase}
\end{eqnarray}
\citep[see][for particular examples]{Brassard95,Wu01}.
In this model, $\omega_{3\pm}$ does not correspond to the frequency of oscillation of an actual mode. Rather, it is produced by frequency mixing due to nonlinear effects, usually in the outer portions of the star's envelope. Since $\omega_1$ and $\omega_2$ correspond to the frequencies of modes in the star, they are often called ``parent'' frequencies and $\omega_3$ is called a ``child'' or ``combination'' frequency.  As already stated, such frequencies, which are observed in $\delta$\,Sct stars, have amplitudes that are typically 3 orders of magnitude smaller than those observed in
KIC\,8054146.

One possible explanation for the large amplitudes is resonant mode coupling. In this scenario there is a damped mode in the star with a frequency very close to the sum of the frequencies of two other linearly unstable modes, i.e., $\omega_3 \approx \omega_1 +\omega_2$; this situation was first studied by \citet{Dziembowski82}. If we take the amplitudes of the parent modes to be $A_1$ and $A_2$, then the steady-state solution of eqn.~2.25 of \citet{Dziembowski82} gives
\begin{equation}
  A_3 = \frac{H}{2\sigma_3\gamma_3 I_3} A_1 A_2.
  \label{amp_coupled}
\end{equation}
Here, $H$ is a coupling coefficient, and $\sigma_3$, $\gamma_3$, and $I_3$ are the dimensionless frequency, damping rate, and inertia of mode 3. We term this the ``mode coupling model''. If the product of $\gamma_3$ and $I_3$ is sufficiently small, then $A_3$ can be much larger than in the simple nonlinear mixing case, and could explain the large amplitudes in this star. With the exception of the potentially larger proportionality factor for the amplitude, we note that the mode coupling model predicts the same relation between the frequencies, phases, and amplitudes of the parent and coupled modes given by eqs.~\ref{freq}, \ref{phase}, and \ref{amp_coupled}. 

Of course, eq.~\ref{amp_coupled} was derived for the case of no rotation, whereas the star KIC 8054146 is rotating at a significant fraction of break-up and is thus non-spherical; this will result in modifications of the mode eigenfunctions and eigenfrequencies \citep{Reese09}.  While difficult to calculate, each mode can still be characterized with a damping rate and a mode inertia. We \emph{posit} that lowest-order nonlinear coupling of modes in such a star could lead to resonant mode coupling as in eq.~5, but with values of $H$, $I_3$, and $\gamma_3$ that are modified from the non-rotating case.

Applying this model (eqs.~3--5) leads to three observational tests. The first test is that the frequencies need to be related as in eq.~\ref{freq}; written in terms of observed frequencies this is $f_{\rm coupled} = f_{\rm parent\, 1} \pm f_{\rm parent\, 2}$. Such an agreement alone is not sufficient to demonstrate mode coupling, since the large number of detected frequencies may lead to accidental agreements.

The other two tests are related to the phase and amplitude constraints, eqs.~\ref{phase} and \ref{amp_coupled}, respectively. For the case in which the phase and amplitude of the modes are constant in time, we are limited in what we can learn. We can measure whether the phase condition (eq.~\ref{phase}) holds for this set of modes, which helps establish that the three modes have a parent/coupled mode relationship, although we cannot tell which are the parent modes and which is the coupled mode.  In the case of amplitudes, it is difficult to test the relationship given by eq.~\ref{amp_coupled} since the value of the coupling constant $H$ and the constants $\gamma_3$ and $I_3$ are not known a priori.

The situation is much better in the case of variable phases and amplitudes. First, instead of providing a single global test of the phase and amplitude relations, each data point provides an independent check. Second, the case of variable amplitudes allows a unique application of the simple mode-coupling model: the amplitude variations of a coupled mode should follow the product of the amplitude changes of the two parent modes, independent of the unknown value of the coupling constant. This allows us to \emph{observationally} determine which are the parent modes and which are the coupled modes.

In order to separately establish that the nine dominant modes are related and to separate these into parent and coupled modes, we have considered all possible combinations and computed phase (O-C) changes and amplitude changes. Furthermore, we also tested a number of possible multiple dependencies. The results were surprisingly unambiguous and clear.

\subsection{Triplet 2}

Due to the qualitative similarity of the amplitude and phase changes of the two triplets (see Figs.~2 and 3), we now compare the amplitudes and phasing for each component of the two triplets.  Our first model assumes that the amplitudes of the modes in Triplet 2 are directly proportional to the amplitudes of the respective modes in Triplet 1; we call this the ``simple scaling model.''  This model is shown in the bottom panel of Fig.~4 and, while hinting at the trend, does not accurately predict the Triplet 2 amplitudes as a function of the Triplet 1 amplitudes.

The situation is considerably improved if we use the mode coupling model in which Triplet 2 is the result of coupling between Triplet 1 and mode f26. In this case, the amplitudes in Triplet 2 are given by a product of the respective amplitudes in Triplet 1 with the amplitude of f26, according to eq.~(\ref{amp_coupled}). The success of this model is shown in the top panel of Fig. 4: the amplitude variations of all three triplet components are now correctly matched. In fact, for the upper component of Triplet 2, the amplitudes are predicted to $\pm 0.002$ ppt on the average! Note that this shows that f26 and Triplet 1 contain the parent modes, while Triplet 2 consists of the coupled modes.

\begin{figure}[!t] 
\centering 
\includegraphics[bb=5 40 540 740, width=\columnwidth, clip]{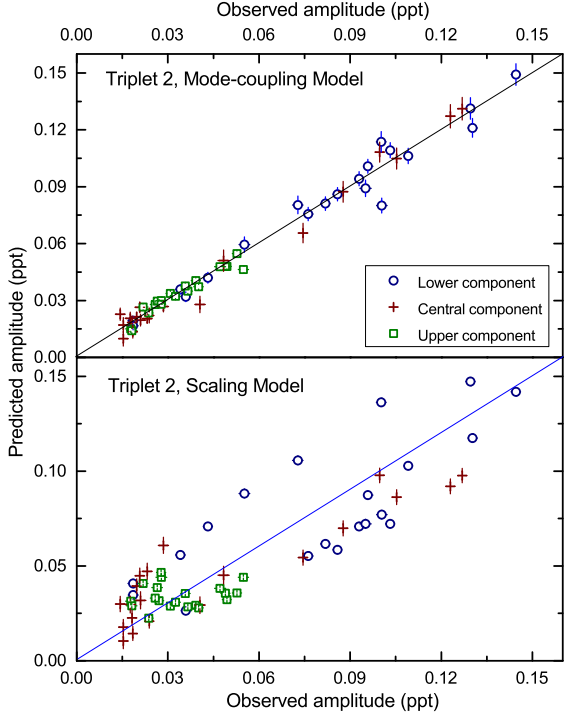} 
\caption{The relationship between the observed and predicted amplitudes for all three components of the frequency Triplet 2. This diagram illustrates that the simple scaling model using the amplitudes of the Triplet 1 components alone are not sufficient (lower panel). However, the mode-coupling model (using the amplitudes of Triplet 1 \emph{and} of mode f26) provides excellent fits (upper panel).}
\end{figure}

\begin{figure}[!htb]
\centering
\includegraphics[bb=60 40 670 570, width=\columnwidth, clip]{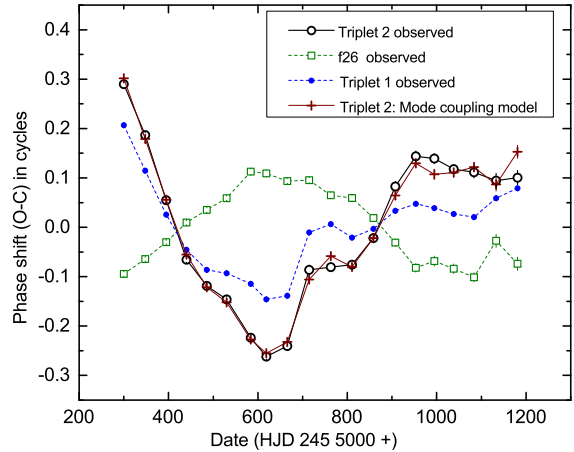}
\caption{(O-C) phase shifts due to changing frequencies for the upper component of Triplet 2. The diagram shows that the observed phase shifts can be modeled by subtracting the phase shifts of f26 from those of Triplet 1, as demanded by the mode-coupling model. Excellent agreement is also found for the other two components of the triplet.}
\end{figure}

The mode-coupling model also requires that the (O-C) phase shifts should match. This excellent agreement is shown in Figure 5 together with two other possible models.

\subsection{The dominant low frequency at 5.86 cycles day$^{-1}$}

There are at least three promising models for the observed frequency peak at 5.8642 cycles day$^{-1}$, all of which are mode-coupling models. The first hypothesis makes this frequency a harmonic of the 2.93 cycles day$^{-1}$ peak(s).
The second and third models involve the outer components of the two triplets. The frequency value of 5.8642 cycles day$^{-1}$ is exactly the difference of the two outer components of either of the two triplets. Consequently, this peak could be a mode coupled with the two outer components of either (or both) Triplet 1 and Triplet 2.

The (O-C) phase shift test supports mode coupling with Triplet 1 (Figure 6) and rejects mode coupling with Triplet 2. This
result is supported by the amplitude test and is illustrated in Figure 7. The figure also rejects the harmonic model, which would explain the 5.86 cycles day$^{-1}$ peak as a consequence of
a nonsinusoidal light-curve shape of the 2.93 cycles day$^{-1}$ mode.

We conclude that the measured phase and amplitude variations of the low frequency at 5.86 cycles day$^{-1}$ are fully compatible with mode coupling of the Triplet 1 parents. This also strengthens our earlier result that Triplet 1 is the parent of a number of different coupled modes, such as Triplet 2.

\begin{figure}[!htb]
\centering
\includegraphics[bb=60 40 670 570, width=\columnwidth, clip]{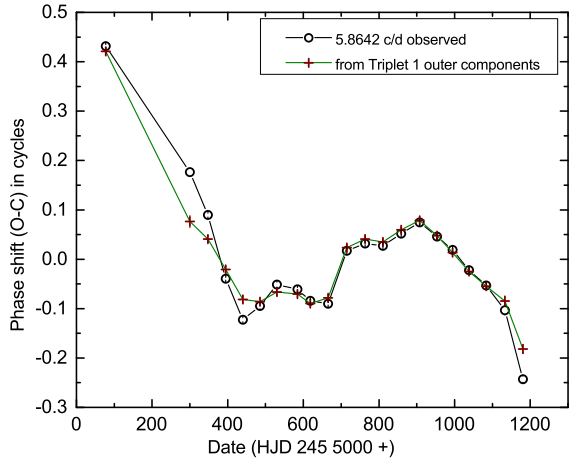}
\caption{(O-C) phase shifts of the low-frequency peak at 5.8642 cycles day$^{-1}$. This diagram shows that the observed phase shifts correspond to the difference of the phase shifts of the outer components of Triplet 1. In order to examine (and reject) a slow phase drift, we have added the results from the short Q2.3 data set (points on left).}
\end{figure}

\begin{figure}[!htb]
\centering
\includegraphics[bb=10 20 730 570, width=\columnwidth, clip]{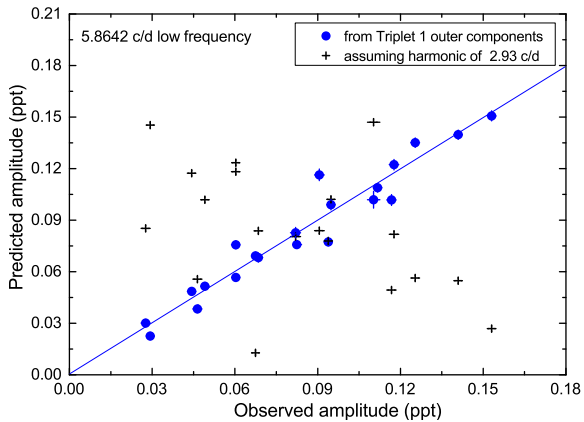}
\caption{Relationship between the observed and predicted amplitude for the low frequency at 5.86 cycles day$^{-1}$. This diagram shows that this frequency is not a harmonic of the 2.93 cycles day$^{-1}$ peak,
but related to the amplitudes of the outer components of the high-frequency Triplet 1.}
\end{figure}

\subsection{The dominant low frequency at 2.93 cycles day$^{-1}$}

\begin{table}
\caption{Amplitudes of the 2.93 cycles day$^{-1}$ frequency}
\begin{tabular}{lc}
\hline
\noalign{\smallskip}
Model & Average residuals (ppt)\\
& Observed - Predicted\\
\noalign{\smallskip}
\hline
\noalign{\smallskip}
Triplet 1 combination& 0.015\\
Triplet 1, f66*f63 & 0.017\\
Triplet 1 central comp.& 0.021\\
Triplet 2 combination & 0.050\\
If 5.86 cycles day$^{-1}$\\
\hspace{5mm}were a harmonic of f3& 0.058\\
\hline
\end{tabular}
\end{table}

In a few amplitude spectra combining data from different quarters (e.g., Q11 to Q14) for increased frequency resolution, we see a small side-peak or asymmetry near 2.933 cycles day$^{-1}$, which is caused by the (smaller) coupling of f63 with f60. Because this peak is uncertain, we have concentrated on the dominant 2.9305 cycles day$^{-1}$ peak.

We have tested five models: (i) mode coupling between all three components of Triplet 1 with a random-walk-type addition of the
amplitudes of f66/f63 and f63/f60, (ii) mode coupling between f66 and f63 of Triplet 1, (iii) amplitude scaling of only the central mode of Triplet 1, since that mode is responsible for most of the amplitude variations, (iv) mode coupling between the three components of Triplet 2, and (v) assuming that 5.86 cycles day$^{-1}$ is a harmonic of 2.93 cycles day$^{-1}$, i.e. scaling the observed amplitudes of the 5.86 cycles day$^{-1}$ peak.

\begin{deluxetable}{lllll}
\tablecolumns{5}
\tablewidth{0pc}
\tablecaption{Summary of observational findings}
\tablehead{
\multicolumn{2}{c}{Identification} & Frequency & Amplitude & Phase\\ 
& & matched by & matched by & matched by}  
\startdata
f34 & Triplet 2 & f60 -- f26 & f26 * f60 & f60 -- f26\\
f37 & Triplet 2 & f63 -- f26 & f26 * f63 & f63 -- f26\\
f40 & Triplet 2 & f66 -- f26 & f26 * f66 & f66 -- f26\\
f6 & low frequency & f66 -- f60 & f66 * f60 & f66 -- f60\\
f3, f3b & low frequency doublet &  f66 -- f63, f63 -- f60 & f66*f63, f60*f63 & ...\\
f86 & Triplet 3 & f60 + f26 & ... & ...\\
f92 & Triplet 3 & f66 + f26 & f26 * f66 & ...
\enddata
\vspace*{-2em}
\tablecomments{Parent frequencies are f26 and f60, f63, f66 (Triplet 1).
The f3/f3b peak has two components from the upper and lower component of Triplet 1, resp.}
\end{deluxetable}

The results for the different quarters (Q5 to Q14) are shown in Table 2. The first two models track the observed variations well, as expected from the mode-coupling model. We can immediately eliminate the Triplet 2 model as well as the
hypothesis that the 5.86 cycles day$^{-1}$ peak is a harmonic of the 2.93 cycles day$^{-1}$ peak.

\subsection{A predicted third triplet}

The mode coupling of Triplet 1 with f26 results in Triplet 2 as well as two dominant low frequencies. This represents the frequency-difference term of the mode-coupling equation (eq.~\ref{freq}).  The mode-coupling equation also predicts the possibility of a frequency-sum term, i.e., an additional triplet at very high frequencies. These components would be found at 86.3855, 89.3189 and 92.2497 cycles day$^{-1}$ (hereafter called f86, f89, and f92).

Two of these predicted modes from the third triplet were detected before with small amplitudes in our analysis of the Q5 to Q10 data, which reported 349 statistically significant frequencies. We detected frequencies of 86.3858 $\pm$ 0.0003 and 92.2497 $\pm$ 0.0002 cycles day$^{-1}$, which are in excellent agreement with the predictions from the present mode-coupling model. The amplitude signal/noise ratios were 4.7 and 7.1, respectively. These values represent a statistically significant detection (Breger et al. 1993). The f89 mode was not seen and is probably hidden in the noise.

The mode-coupling model also predicts the shape of the amplitude variability of the Triplet 3 modes. Only f92 has an amplitude large enough to examine its variability on a quarterly basis. We find a variation from 1.6 to 5.4 ppm from Q5 to Q14. This variation accurately tracks the predicted changes of amplitude. We find an average deviation between the predicted and observed amplitude of only $\pm$ 0.4 ppm per quarter. (At first sight, such extremely small deviations may be appear unrealistic, since the formal uncertainty in the quarterly amplitude solution is 1 ppm. However, the calculation of the formal uncertainties assumes white noise. The real noise at such high frequencies is much lower.)

Thus, not only did the mode coupling model predict the existence of these additional frequencies, it correctly predicted the amplitude variation of the only component of the new third triplet for which this test could be made. We summarize our results in Table 3.

\section{Alternate explanations}

\subsection{Combination-mode hypothesis}

Combination modes are usually observed in A/F stars, i.e., two simultaneously excited modes f$_i$ and f$_j$ lead to (f$_i$ $\pm$ f$_j$), observed at small amplitudes. (Note that these frequencies, commonly called combination modes, may not be normal modes, but nonlinearities in the star.) A typical example is the star 44 Tau, in which combinations of different $\ell$ values are detected (Breger \& Lenz 2008). To quantify the size of these combination amplitudes we introduce 
the parameter $\mu$, defined as follows:
\begin{equation}
A_{\rm comb} = \mu \cdot A_i \cdot A_j,
\end{equation}
where $A_i$ and $A_j$ are the amplitudes of the parent modes (in fractions).  Typical values of $\mu$ are found to be about 4, which has also been seen in other stars ($\mu$ is observed to lie within a factor of 2 of this value). Furthermore, the amplitudes of the (f$_i$ - f$_j$) combinations are generally slightly smaller than those of (f$_i$ + f$_j$).
We have already noted earlier that KIC\,8054146 exhibits many of these typical combination frequencies and their small amplitudes are typical for $\delta$ Scuti stars.

We can now compare the values above to those found for the coupled modes in KIC\,8054146. There is a wide range of $\mu$ values ranging from 7000 to 60000 for Triplet 2, and 11000 for the 5.86 cycles day$^{-1}$ low frequency. Even for the small-amplitude upper component of Triplet 3, we find a value near 700.  We conclude that the observed coupling coefficients in this star are three to four orders of magnitude larger than those observed in other $\delta$~Scuti stars. This makes it unlikely that these frequencies are produced by the same mechanism that generates combination frequencies in other $\delta$ Scuti stars.

The only way to rescue the combination hypothesis involves fortuitous geometry.  In photometry we measure the integrated light across the visible disk. For nonradial modes, the observed amplitudes of pulsation suffer from cancellation effects due to this integration.  If a star is viewed nearly equator on, the measured amplitude of a nonradial $\ell$ = 1, $m$ = 0 mode, for example, will be severely decreased. In the present discussion, we eliminated the combination-mode hypothesis because the observed amplitudes of the coupled modes were too large by about four orders of magnitude. Could it be that instead the observed amplitudes of one or more parents are too small by these huge factors, i.e., that the star is seen essentially exactly equator-on? Let us assume that one of the parent modes actually has an amplitude larger by four orders of magnitude. Inspection of Fig. 2 shows that in this case, the intrinsic amplitude of the parent mode would be much larger than any known amplitude for a rapidly rotating $\delta$ Scuti star. Moreover, at least one of the parent modes of each of the coupled mode frequencies reported in this paper would also have to suffer from such geometric cancellation. 

Consequently, we find the combination-mode hypothesis as well as the cancellation argument very improbable.

\subsection{Asymptotic pulsation}

The nearly equidistant spacing of the parent triplet identified in section~3 must be a result of the underlying frequency spectrum in the star. And, if the higher frequency child triplet is due to resonant mode coupling, then the star must have eigenfrequencies very close to these values as well, i.e., this second equally-spaced triplet must also exist in the eigenfrequency spectrum of the star. 

For a non-rotating star, the frequency spectrum of p~modes in the asymptotic limit is given by
\begin{equation}
f_{n \ell} \simeq \left(n + \frac{\ell}{2} +\frac{1}{4} + \alpha\right) \Delta \nu,
\label{asymp}
\end{equation}
where
\begin{equation}
\Delta \nu = \left[2 \int_0^R \frac{dr}{c}\right]^{-1} 
\end{equation}
is the inverse sound crossing time and we have ignored the small separation \citep{Unno89}.  
Thus, modes differing by $\ell=\pm 1$ or $n=\pm 1$ would appear nearly equally spaced.

As it turns out, a modified form of equation~\ref{asymp} exists for the case of rapid rotation:
\begin{equation}
\omega_{n,\ell,m} = \tilde{\Delta}_n \tilde{n} + \tilde{\Delta}_\ell \tilde{\ell} + \tilde{\Delta}_m |m| + \alpha^{\pm },
\end{equation}
where $\tilde{n} = 2 n+\epsilon$, $\tilde{\ell} = (\ell -|m| - \epsilon)/2$, and the coefficients $\tilde{\Delta}_n$, $\tilde{\Delta}_\ell$, and $\tilde{\Delta}_m$ all depend on the structure of the star. \citep{Reese08}. For this case we again see that evenly spaced modes can be produced by $\tilde{\ell}=\pm 1$ or $\tilde{n}=\pm 1$ modes. Thus, evenly spaced modes can be a natural consequence of rapid rotation in this star.

\subsection{Rotational modulation of two p-modes}

Small amplitude modulations of p-modes, caused by rotation, have been detected before and may be very common even in normal A and F stars, e. g., in KIC\,9700322 (Breger et al. 2011). The 2.93 cycles day$^{-1}$ peak would be the rotation frequency, which would agree with the known high measured stellar rotation (Paper I), while 5.86 cycles day$^{-1}$ would be the 2f harmonic. The problem of rotational frequency peaks with f and 2f of similar amplitudes has been noted before (Balona 2011) and awaits explanations. The upper and lower components of Triplets 1 and 2 would be caused by the rotational amplitude modulation of the central component.

This attractive hypothesis fails because the side peaks associated with the amplitude modulation are not symmetric and at most times are larger than the actual pulsation itself.

\section{Is KIC 8054146 unique?}

It might be argued that KIC 8054146 is unusual or unique due to its very high stellar rotation of $v\sin i$ = 300 km s$^{-1}$ (Paper I) and almost equidistant frequency triplets. However, many of the properties shown by this star are shared by other stars of spectral type A/F.
In this discussion, let us ignore ``normal'' combinations, $nf_i$ - $mf_j$, where $f_i$ and $f_j$ are the frequencies of two pulsation modes with the combination frequencies having much smaller amplitudes than $f_1$ and $f_2$. Related frequency peaks of similar amplitudes have been seen before in a number of A/F stars, even between the low-frequency and high-frequency regions.
The present results on KIC 8054146 are unusual because the parent and child (coupled) modes could be uniquely identified: any two of four identified parent frequencies (Triplet 1 and 25.95 cycles day$^{-1}$) can correctly predict the variable amplitudes of a number of other frequencies.

Let us compare KIC 8054146 with two other A/F stars. Our unpublished analysis of KIC 9664869 also shows equidistant frequency spacings and corresponding relationships between low and high frequencies with relatively high amplitudes. However, the available data are insufficient to use the amplitude variations to separate the parent from the child modes.

Balona et al. (2013) studied the roAp star KIC 8677585.  As in KIC 8054146, the value of the dominant low frequency corresponds to the spacing of some high-frequency modes. Furthermore, it shares the amplitude and frequency variations of a high-frequency mode at 140.1 cycles day$^{-1}$, suggestive of nonlinear interactions. Note that in KIC 8054146, such a similarity with the behavior of a single
mode is only superficial and the variations of the low-frequency peak are matched accurately only by the coupling of $\it two$ parent modes. However, the data (and amplitudes) for KIC 8677585 are insufficient to test mode coupling between two modes and identify the parents. Until more data of such other intriguing pulsators become available, the main result of the present paper cannot be fully tested in other stars.

The amplitude variability on a time scale of years found in KIC 8054146 is not unique either for nonradial or even for some radial modes. An example is the star 4 CVn (Breger 2009), for which systematic changes as a function of the azimuthal number, $m$, were detected over 40 years.

Let us now turn to the relatively high frequencies of the parent Triplet 1 (60, 63 and 66 cycles day$^{-1}$). Are they unusually high for a $\delta$ Sct star? They certainly do not fit the
``classical'' picture of stellar excitation shown by the HADS, the slowly rotating high-amplitude-Delta Scuti stars. In this picture, pulsation is excited to high amplitude in a narrow range of frequencies corresponding to low-order radial modes.
However, using ground-based telescopes, Rodr\'{i}guez et al. (2007) already showed that in the HADS star BL Cam, low-amplitude nonradial modes were detected up to 79 cycles day$^{-1}$. These were explained by Breger et al. (2009) as nonradial modes trapped in the envelope of the star.

The accurate, recent observations by $\it Kepler$ and $\it CoRoT$ have further supported the detection of high-frequency oscillations in $\delta$ Sct stars, e.g., see the catalog by Uytterhoeven et al. (2011). We note here that some of the high frequencies listed in the catalog are low-amplitude combination frequencies, rather than independent pulsation modes. Also, some frequency peaks may actually be low frequencies in the stellar frame of reference, shifted to high frequencies in the observer's frame of reference due to rotational splitting (Breger et al. 2013). Consequently, the detected high frequencies should be individually examined  for each star. Nevertheless, based on the recent observations of a multitude of $\delta$ Sct stars, the frequencies of the Triplet 1 near 63 cycles day$^{-1}$, which is the main mode-coupling parent in KIC 8054146, is not unusual. Furthermore, present models of pulsation instability as a function of frequency confirm that
high frequencies can indeed be excited (Daszynska-Daszkiewicz et al. 2005, 2006), but the results depend critically on the treatment of convection.

We conclude that while KIC 8054146 may be the first $\delta$ Sct star for which mode coupling and the identification of the parents and the children are reported, many of the star's properties are shared by other $\delta$ Sct stars.
At this stage, considerably more data are required in order to test the mode-coupling hypothesis for these stars.

\section{Discussion}
 
For KIC 8054146 we identify coupled modes and can uniquely distinguish which are the parent and child modes. This is made possible by the fact that for the simplest, lowest-order mode coupling the amplitudes of the coupled modes are given by the product of the amplitudes of the parents. We speculate that the simple mode-coupling model is applicable due to the relatively small amplitudes and therefore lowest-order nonlinear interaction of the modes involved. Consequently, the present study is only made possible by the excellent time coverage and precision of the data from the {\it Kepler} spacecraft. We note that the generally unknown mode-coupling constants do not present a problem for KIC 8054146 due to the large amount of amplitude variability over the three years of data; a single short stretch of data allows us to derive the mode-coupling coefficients, which subsequent data groups show to be constant.

The state of nonlinear or combination frequencies detected in KIC\,8054146 is the following:
\begin{enumerate}
 \item We detect ``normal'' nonlinearities in this star; these are combination frequencies for which the amplitudes are related by $A_3\sim  \mu * A_1 * A_2$, where $\mu$ is $\sim 4$. These small nonlinearities are commonly seen in most if not all Delta Scuti stars. 

 \item KIC 8054146 also possesses larger-amplitude nonlinear signals, and we are able to use the measured amplitude and phase variations of these signals to determine which modes are the ``parent'' modes and which are the ``child'' modes. The derived value of $\mu$ relating the  amplitudes of child and parent modes is $\mu \sim1000$ to 10,000 or more. 
\end{enumerate}

A generic nonlinear model (such as the nonlinear relation between temperature and flux at the surface) is not able to simultaneously produce both these ``small'' and ``large'' nonlinearities. On the other hand, if a subset of the combination frequencies lie close to eigenfrequencies in the star, then these modes and only these modes can be resonantly driven to larger amplitudes. While this is far from the last word on the subject, the data are fully consistent with this ``resonant mode coupling'' scenario.

 In particular, the present investigation showed that in KIC 8054146, two dominant low frequencies are actually coupled modes with amplitudes larger or equal to those of the dominant parents in the higher-frequency domain. The low-frequency region of these so-called hybrid pulsators contains a large number of linearly driven gravity modes, as was confirmed by Chapellier et al. (2012) from searching for equidistant period spacings in ID 105733033. Furthermore, in a number of hybrid pulsators, some frequency values in the gravity-mode and pressure-mode domains are numerically not independent of each other. Given the existence of these gravity modes, we speculate that these could be the ``child'' modes that are resonantly excited to larger amplitudes. A confirmation of this resonance explanation would be provided if additional data could reveal that these particular low frequencies are part of a sequence of low-frequency modes equidistantly spaced in period.  

\acknowledgments

It is a pleasure to thank E. L. Robinson for helpful discussions. This investigation has been supported by the Austrian Fonds zur F\"{o}rderung der wissenschaftlichen Forschung through project P\/21830-N16, by the Norman Hackerman Advanced Research Program under grant 003658-0252-2009, the National Science Foundation under grants AST-0909107 and AST-1312983, and NASA under grant NNX12AC96G. The authors also wish to thank the {\it Kepler} team for their outstanding efforts which have made these results possible.  Funding for the {\it Kepler} mission is provided by NASA's Science Mission Directorate.


\begin{thebibliography}{}

\bibitem[Balona(2011)]{Balona11} Balona, L. A.\ 2011, \mnras, 415, 1691
\bibitem[Balona(2013)]{} Balona, L. A., Catanzaro, G., Crause, L., et al. 2013, \mnras, 432, 2808
\bibitem[Brassard et al.(1995)]{Brassard95} Brassard, P., Fontaine, G., \& Wesemael, F.\ 1995, \apjs, 96, 545
\bibitem[Breger (2009)]{Breger09} Breger, M.\ 2009, AIP Conf. Proc., 1170, 410
\bibitem[Breger et al.(1993)]{Breger93} Breger, M., Stich, J., Garrido, R., et al. 1993, A\&A, 271, 482
\bibitem[Breger et al.(2011)]{Breger11} Breger, M., Balona, L., Lenz, P., et al.\ 2011, \mnras, 414, 1721 
\bibitem[Breger \& Lenz(2008)]{Breger08} Breger, M., \& Lenz, P.\ 2008, \aap, 488, 643
\bibitem[Breger et al.(2012)]{Breger12} Breger, M., Fossati, L., Balona, L., et al.\ 2012, \apj, 759:62 (Paper I)
\bibitem[Breger et al.(2009)]{Breger09} Breger, M., Lenz, P., \& Pamyatnykh, A. A. 2009, \mnras, 396, 291
\bibitem[Breger et al.(2013)]{Breger13} Breger, M., Lenz, P., \& Pamyatnykh, A. A. 2013, \apj, 773:56
\bibitem[Chapellier et al.(2012)]{C12} Chapellier, E., Mathias, P., Weiss, W., Le Contel, D., \& Debosscher, J. 2012, A\&A, 540, A117
\bibitem[Daszynska-Daszkiewicz et al.(2005)]{Daszynska05} Daszy\'nska-Daszkiewicz, J., Dziembowski, W.~A., Pamyatnykh, A.~A., et al. 2005, A\&A, 438, 653
\bibitem[Daszynska-Daszkiewicz et al.(2006)]{Daszynska06} Daszy\'nska-Daszkiewicz, J., Dziembowski, W.~A., \& Pamyatnykh, A.~A.\ 2006, MmSAI, 77, 113
\bibitem[Dziembowski(1982)]{Dziembowski82} Dziembowski, W.\ 1982, \actaa, 32, 147 
\bibitem[Monnier et al.(2010)]{Monnier10} Monnier, J.~D., Townsend, R.~H.~D., Che, X., et al.\ 2010, \apj, 725, 1192
\bibitem[Reese et al.(2008)]{Reese08} Reese, D., Ligni{\`e}res, F., \& Rieutord, M.\ 2008, \aap, 481, 449 
\bibitem[Reese et al.(2009)]{Reese09} Reese, D.~R., MacGregor, K.~B., Jackson, S., Skumanich, A., \& Metcalfe, T.~S.\ 2009, \aap, 506, 189 
\bibitem[Rodriguez et al.(2007)]{Rodriguez7}Rodr\'{i}guez, E., Fauvaud, Farrell, J. A., et al.\ 2007, A\&A, 471, 255
\bibitem[Unno et al.(1989)]{Unno89} Unno, W., Osaki, Y., Ando, H., Saio, H., 
\& Shibahashi, H.\ 1989, Nonradial oscillations of stars, Tokyo: University of Tokyo Press, 1989, 2nd ed.
\bibitem[Uytterhoeven et al.(2011)]{Uytterhoeven11} Uytterhoeven, K., Moya, A., Grigahc{\`e}ne, A., et al.\ 2011, \aap, 534, A125 
\bibitem[Wu(2001)]{Wu01} Wu, Y.\ 2001, \mnras, 323, 248 
\end{thebibliography}
\end{document}